\documentclass[onecolumn,amsmath,amssymb,prl]{revtex4}

 \usepackage{graphicx} % Include figure files
 \usepackage{dcolumn}  % Align table columns on decimal point
 \usepackage{bm}       % bold math

\begin{document}
\draft

\title{{\it Ab Initio} determination of Cu $3d$ orbital energies in layered copper oxides}

\author{Liviu Hozoi$^{1,\dagger}$, Liudmila Siurakshina$^{1,2}$, Peter Fulde$^3$, Jeroen van den Brink$^1$}
\affiliation{
$^1$Institute for Theoretical Solid State Physics, IFW Dresden, Helmholtzstr.~20, 01069 Dresden, Germany\\
$^2$Laboratory of Information Technologies, Joint Institute for Nuclear Research, 141980 Dubna, Russia \\
$^3$Max-Planck-Institut f\"{u}r Physik komplexer Systeme, N\"{o}thnitzer Str.~38, 01187 Dresden, Germany\\
$^{\dagger}$Correspondence and requests for materials should be addressed to L.H.~(l.hozoi@ifw-dresden.de)}

\maketitle

%% \section{Abstract}

\textsl{
It has long been argued that the minimal model to describe the low-energy physics of
the high $T_{\rm c}$ superconducting cuprates must include copper states of other
symmetries besides the canonical $3d_{x^2-y^2}$ one, in particular the $3d_{z^2}$ orbital.
Experimental and theoretical estimates of the energy splitting of these states vary widely.
With a novel {\it ab initio} quantum chemical computational scheme we determine these
energies for a range of copper-oxides and -oxyclorides, determine trends with the apical 
Cu--ligand distances and find excellent agreement with recent Resonant Inelastic X-ray
Scattering measurements, available for La$_2$CuO$_4$, Sr$_2$CuO$_2$Cl$_2$, and CaCuO$_2$.
}

%% \newpage

 \ 

 \ 

 \ 

%% \section{Introduction}

It is generally accepted that the low-energy physics of the layered Cu oxide
compounds in their normal state is reasonably well described by models which incorporate
the ``in-plane'' Cu $3d_{x^2-y^2}$ and O $2p_x$/$2p_y$ orbitals.
However, the energy window over which such models provide a qualitatively correct picture
is a matter of active research.
One additional ingredient which is often invoked is the Cu $3d_{z^2}$ orbital, perpendicular
onto the CuO$_2$ layers, and the apical O $2p_z$ functions having $\sigma$-type
overlap with the Cu $3d_{z^2}$.
Recent multi-orbital calculations using dynamical mean-field theory \cite{CuO2_z2_weber10}
show indeed that some of the features of the optical, x-ray absorption, and photoemission
spectra can be better reproduced when the Cu $3d_{z^2}$ orbitals are explicitly included
in the many-body treatment. 
At finite doping, the inclusion of the Cu $3d_{z^2}$ functions makes a difference even
for the low-energy states close to the Fermi level \cite{CuO2_z2_weber10}. 

The off-diagonal coupling between states of $x^2\!-\!y^2$ and $z^2$ symmetry was actually
found to substantially affect the dispersion of the low-energy bands and the shape of
the Fermi surface in earlier semiphenomenological models \cite{CuO2_z2_eskes91,CuO2_z2_feiner96},
density-functional calculations \cite{CuO2_z2_oka01}, and quantum chemical studies
\cite{CuO2_QPs_hozoi08}.
Moreover, Ohta {\it et al.} \cite{CuO2_z2Tc_ohta91} and recently Sakakibara {\it et al.}
\cite{CuO2_z2Tc_sakaki10} suggested that a direct relation exists between the magnitude
of $T_{\rm c}$ and the size of the $d_{x^2-y^2}$--$d_{z^2}$ splitting. 
The splittings within the Cu $3d$ shell are also relevant to excitonic models for pairing
and high-$T_{\rm c}$ superconductivity \cite{CuO2_ddTc_cox89,CuO2_ddTc_little07}. 
Even if  the importance of the Cu $3d_{z^2}$ state is stressed in this considerable body of work,  the 
actual experimental and theoretical estimates of the energy of this state vary widely.

Sharp features at about 0.4 eV in early optical measurements on La$_2$CuO$_4$ and Sr$_2$CuO$_2$Cl$_2$
were initially assigned to crystal-field Cu $d_{x^2-y^2}$ to $d_{z^2}$ charge excitations
\cite{CuO2_dd_perkins93}.
A different interpretation in terms of magnetic excitations was proposed by Lorenzana and Sawatzky
\cite{CuO2_J_lorenzana95} and latter on confirmed by analysis of the resonant inelastic x-ray
scattering (RIXS) spectra at the Cu K and $L_3$-edge \cite{CuO2_rixsJ_Hill08,CuO2_rixsJ_09}.
The RIXS experiments also show that in La$_2$CuO$_4$ and Sr$_2$CuO$_2$Cl$_2$ the Cu
$d_{x^2-y^2}$ to $d_{z^2}$ transitions occur at 1.5--2.0 eV \cite{CuO2_rixsdd_10}, which is
substantially larger than the outcome of earlier wavefunction-based quantum chemical calculations,
1.0--1.2 eV \cite{CuO2_dd_coen00}, or density-functional estimates, 0.9 eV
\cite{CuO2_z2Tc_sakaki10}.

With the aim to settle this point we employ a recently developed {\it ab initio} quantum chemical 
computational scheme to extract the splittings within the Cu $3d$ shell in several layered
copper oxides. 
Excellent agreement is found for La$_2$CuO$_4$, Sr$_2$CuO$_2$Cl$_2$, and CaCuO$_2$
with recent RIXS measurements \cite{CuO2_rixsdd_10}.
%% Deviations as large as 0.5 eV in earlier quantum chemical investigations \cite{CuO2_dd_coen00}
%% are assigned to the less precise representation in those studies of the surroundings 
%% of the Cu ion for which the $d$--$d$ transitions are calculated. 
Further, the $d_{x^2-y^2}$ to $d_{z^2}$ excitation energies computed here for La$_2$CuO$_4$, 
YBa$_2$Cu$_3$O$_6$, and HgBa$_2$CuO$_4$ are relevant to models which attempt to establish
a direct relation between the relative energy of the out-of-plane $d_{z^2}$ level
and the critical temperature $T_{\rm c}$ \cite{CuO2_z2Tc_sakaki10}.
In particular, the large difference between the critical superconducting temperatures of
La$_2$CuO$_4$ and HgBa$_2$CuO$_4$ was directly attributed to a large difference between
the $d_{x^2-y^2}$ to $d_{z^2}$ excitation energies \cite{CuO2_z2Tc_sakaki10}.

\section{Results}

To study bound, excitoniclike states such as the $d$--$d$ charge excitations in copper
oxides, we rely on real-space {\it ab initio} methods.
In the spirit of modern multi-scale electronic-structure approaches, we describe a given
region around a central Cu site by advanced quantum chemical many-body techniques while
the remaining part of the solid is modeled at the Hartree-Fock level.
The complete-active-space self-consistent-field (CASSCF) method was used to generate
multireference wavefunctions for further configuration-interaction (CI) calculations
\cite{QC_book_00}.
In the CASSCF scheme, a full CI is carried out within a limited set of ``active'' orbitals,
i.e., all possible occupations are allowed for those active orbitals.
The active orbital set includes in our study all $3d$ functions at the central Cu site 
and the $3d_{x^2-y^2}$ functions of the Cu nearest neighbor (NN) ions.
Strong correlations among the $3d$ electrons are thus accurately described.
The final CI calculations incorporate all single and double excitations from the Cu $3s$,$3p$,$3d$
and O $2p$ orbitals on a given CuO$_4$ plaquette and from the $3d_{x^2-y^2}$ orbitals of the
Cu NN's.
Such a CI treatment is referred to as SDCI.
The CASSCF and SDCI investigations were performed with the {\sc molpro} quantum chemical software 
\cite{molpro_2006}.

Both SDCI and RIXS results for the Cu $d$-level splittings are listed in Table I.
The relative energies of the peaks observed between 1 and 3 eV in the Cu $L_3$-edge RIXS
spectra \cite{CuO2_rixsdd_10} are the sum of a crystal-field contribution, i.e., an
on-site crystal-field splitting $E_{\rm cf}$, and a magnetic term $\Delta E_{\rm mgn}$.
The quantum chemical calculations have been performed to extract $E_{\rm cf}$ for
a ferromagnetic (FM) arrangement of the Cu $d$ spins.
A SDCI treatment for an antiferromagnetic (AF) alignment of the Cu spins in the
embedded cluster of five Cu sites is computationally not feasible (see Methods for details).
$\Delta E_{\rm mgn}$ accounts for AF order in the ground-state configuration of the Heisenberg
antiferromagnet and is determined as follows.
First the value for the NN exchange coupling constant $J$ is computed by considering
an embedded cluster consisting of two CuO$_4$ plaquettes.
For CaCuO$_2$, for example, we find $J\!=\!0.13$\,eV, in good agreement with the
theoretical results reported in Ref.~\cite{footnote_J} and with values from experimental
data \cite{CuO2_J_lorenzana95,CuO2_rixsJ_09,CuO2_rixsdd_10}.
With this value of $J$ in hand we return to the cluster with five Cu sites and flip the spin
of the central Cu ion.
This corresponds to an energy increase $\Delta E = z J/2 = 2J$, where
$z\!=\!4$ is the number of NN's and we neglect the quantum fluctuations.
For the crystal-field excited states, the superexchange with the NN Cu $d_{x^2-y^2}$ spins is
much weaker for a hole excited into the $d_{z^2}$ orbital and zero by symmetry for a hole into
a $t_{2g}$ orbital.
This contribution due to intersite $d_{z^2}$--$d_{x^2-y^2}$ superexchange,
$\Delta E' = 2J'$, is not included either in the quantum chemical calculations
but in a first approximation we can neglect the weak intersite AF interaction $J'$ involving
a $d_{z^2}$ hole.
From overlap considerations, $J'$ is only a small fraction of the ground-state superexchange
$J$.
For a meaningful comparison between the SDCI and RIXS data, we subtracted in Table I
from the relative RIXS energies reported in Ref.~\cite{CuO2_rixsdd_10} the term
$\Delta E_{\rm mgn} = \Delta E - \Delta E' \approx 2J$ 
representing the magnetic stabilization of the ground-state configuration with respect
to the crystal-field excited states.
Since $J\!\approx\!0.13$ eV, $\Delta E_{\rm mgn}\!\approx\!0.26$.

The agreement between our SDCI excitation energies and the results from RIXS is remarkable.
As shown in Table I, the differences between the SDCI and RIXS energies are not larger
than 0.15 eV.
The only exception is the splitting between the $x^2\!-\!y^2$ and $xz$/$yz$ levels in
CaCuO$_2$, where the SDCI value is 0.3 eV larger than in the RIXS measurements.
That an accurate description of neighbors beyond the first ligand coordination shell
is crucial is clear from the comparison between our and earlier quantum chemical data.
In the calculations described in Ref.~\cite{CuO2_dd_coen00}, only one CuO$_6$ octahedron
or one CuO$_5$ pyramid was treated at the all-electron level.
Farther neighbors were described by either atomic model potentials or point charges.
Deviations of 0.4 and 0.6 eV (up to $50\%$) for the $z^2$ levels in La$_2$CuO$_4$ and 
Sr$_2$CuO$_2$Cl$_2$ \cite{CuO2_dd_coen00}, for example, are mainly due to such 
approximations in the modeling of the nearby surroundings.
The $d$-level splittings depend after all on the charge distribution at the NN ligand
sites.
The latter is obviously sensitive to the manner in which other species in the immediate
neighborhood are modeled. 
The quality of the results reported here is directly related to the size of the
clusters, i.e., five CuO$_4$ plaquettes, all apical ligands plus the NN closed-shell metal
ions of the central polyehdron.

Superconductivity has not been observed in Sr$_2$CuO$_2$Cl$_2$ and CaCuO$_2$.
The $d$-level splittings for three representative cuprate superconductors, i.e., La$_2$CuO$_4$,
HgBa$_2$CuO$_4$, and YBa$_2$Cu$_3$O$_6$ are listed in Table II.
The maximum $T_{\rm c}$'s achieved by doping in these three materials are 35, 95, and 50 K,
respectively.
For the YBa$_2$Cu$_3$O$_6$ compound, we here refer to the maximum $T_{\rm c}$ which can
be achieved by Ca doping \cite{str_YBCO6_mccarron89}.
The large difference between the critical temperatures in La$_2$CuO$_4$ and HgBa$_2$CuO$_4$ 
was assigned in Ref.~\cite{CuO2_z2Tc_sakaki10} to a large difference between the relative
energies of the $z^2$ states in the two materials.
The density-functional results for the splittings between the $x^2\!-\!y^2$ and $z^2$ levels
in La$_2$CuO$_4$ and HgBa$_2$CuO$_4$ are 0.91 and 2.19 eV, respectively \cite{CuO2_z2Tc_sakaki10}.
RIXS data are not available for HgBa$_2$CuO$_4$ and independent estimates for the energy
separation between the $x^2\!-\!y^2$ and $z^2$ states are therefore desirable.
While we find a rather similar value for HgBa$_2$CuO$_4$, of 2.09 eV, the quantum chemical and
RIXS results \cite{CuO2_rixsdd_10} for La$_2$CuO$_4$ are substantially larger, about 1.4 eV.
This makes the difference between the $d$-level splittings in the above mentioned
compounds less spectacular, i.e., $E_{z^2}^{\mathrm{HBCO}} -  E_{z^2}^{\mathrm{LCO}}$
is reduced from 1.3 eV in Ref.~\cite{CuO2_z2Tc_sakaki10} to 0.7 eV in the present study,
which suggests that the model constructed and the conclusions drawn
in Ref.~\cite{CuO2_z2Tc_sakaki10} at least require extra analysis.

The distance between the Cu and apical ligand sites increases from 2.40 \AA  \ in
La$_2$CuO$_4$ \cite{str_LCO_longo73} to 2.78 \AA  \ in HgBa$_2$CuO$_4$
\cite{str_HBCO_wagner93}.
The effect of this growth of the apical Cu--O bond length on the relative energy of the 
$z^2$ hole state can be understood by using simple electrostatic arguments:
when the negative apical ions are closer to the Cu site, less energy is needed to promote
the Cu $3d$ hole into the $z^2$ orbital pointing toward those apical ligands.
For HgBa$_2$CuO$_4$, the lowest crystal-field excitation is therefore to the $xy$ level 
and requires about 1.3 eV, see Table II, while the $z^2$ and $xz$/$yz$ levels are nearly 
degenerate and more than 0.5 eV higher in energy.
On the other hand, in La$_2$CuO$_4$ the lowest crystal-field excitation is to the $z^2$ 
orbital, see Table I.
Our results also reproduce the near degeneracy between the $z^2$ and $xy$ levels in
La$_2$CuO$_4$, as found in the RIXS experiments.
In CaCuO$_2$, there are no apical ligands.
The splitting between the $x^2\!-\!y^2$ and $z^2$ levels is therefore the largest for
CaCuO$_2$, about 2.4 eV, see Table I.

\section{Discussion}

The parameter that plays the major role in determining the size of the $d$-level splittings
in layered cuprates is clearly the apical Cu--ligand distance.
There are, however, few other factors which come into play such as the number and nature
of the apical ligands, the in-plane Cu--O bond lenghts, buckling of the CuO$_2$ planes,
and the configuration of the farther surroundings.
Trends concerning the relative energy of the $z^2$ hole state in different cuprates are
illustrated in Fig.~1, which includes data for systems having one apical O site
(YBa$_2$Cu$_3$O$_6$), two apical O's (La$_2$CuO$_4$, HgBa$_2$CuO$_4$), two apical Cl
ions (Ca$_2$CuO$_2$Cl$_2$, Sr$_2$CuO$_2$Cl$_2$) or no apical ligand (CaCuO$_2$).
The apical Cu--O distances in La$_2$CuO$_4$ and YBa$_2$Cu$_3$O$_6$, for example, are nearly
the same, 2.40 vs.~2.45 \AA  \ \cite{str_YBCO6_mccarron89,str_LCO_longo73,str_YBCO_katano87}.
In YBa$_2$Cu$_3$O$_6$, however, there is a single apical O.
For this reason the $z^2$ hole state is somewhat destabilized in YBa$_2$Cu$_3$O$_6$ and
lies above the $xy$ hole configuration, see Table II.
Yet since the Cu ion is shifted towards the apical ion, out of the basal O plane, the
$x^2\!-\!y^2$ hole state is also destabilized such that the splitting between the 
$x^2\!-\!y^2$ and $z^2$ levels is finally close to the value found in La$_2$CuO$_4$. 
Further, the apical Cu--ligand distances are slightly larger in Sr$_2$CuO$_2$Cl$_2$ as
compared to HgBa$_2$CuO$_4$, 2.86 vs.~2.78 \AA  , respectively.
The apical ions also have a smaller effective charge in Sr$_2$CuO$_2$Cl$_2$, which
should lead to a larger relative energy of the $z^2$ hole state in Sr$_2$CuO$_2$Cl$_2$
as compared to HgBa$_2$CuO$_4$. 
The fact that the relative energy of the $z^2$ hole state is actually larger in
HgBa$_2$CuO$_4$, see Fig.~1, must be related to the smaller in-plane Cu--O distances
in HgBa$_2$CuO$_4$, 1.94 in HgBa$_2$CuO$_4$ vs.~1.99 \AA \ in Sr$_2$CuO$_2$Cl$_2$,
which stabilizes the ground-state $x^2\!-\!y^2$ hole configuration in the former compound,
and to farther structural details.
From Ca$_2$CuO$_2$Cl$_2$ to Sr$_2$CuO$_2$Cl$_2$, the Cu--Cl separation increases from
2.75 to 2.86 \AA \ \cite{str_CCOC_argyriou95,str_SCOC_miller90} and the energy of the
$z^2$ level from 1.37 to 1.75 eV.

In contrast to the $z^2$ orbitals, the relative energies of the $xy$ levels display much
smaller variations, in an interval of 1.2--1.5 eV, see Tables I and II.
Substantially smaller are also the variations computed for the $xz$/$yz$ levels, in an 
energy window between 1.6 and 2.0 eV.

To summarize, we employ state of the art quantum chemical methods to investigate the
Cu $3d$ electronic structure of layered Cu oxides.
Multiconfiguration and multireference configuration-interaction calculations are carried out on finite
clusters including five CuO$_4$ plaquettes plus additional apical ligand and closed-shell
metal ion NN's.
The localized Wannier functions attached to these atomic sites are obtained from prior
Hartree-Fock computations for the periodic system.
Excellent agreement is found between our theoretical results and recent Cu $L_3$-edge
RIXS data for La$_2$CuO$_4$, Sr$_2$CuO$_2$Cl$_2$, and CaCuO$_2$. 
RIXS is a novel experimental tool to investigate both magnetic and charge excitations with
high resolution and accuracy.
Our computational scheme and present results indicate a promising route for the modeling 
and reliable interpretation of RIXS spectra in correlated $3d$-metal compounds.
A next step along this path is the computation of transition probabilities and intensities
at the {\it ab initio} level, which requires the explicit calculation of the intermediate
Cu $3p$ core hole wavefunctions.

Further, the excitation energies computed here for La$_2$CuO$_4$, YBa$_2$Cu$_3$O$_6$, and
HgBa$_2$CuO$_4$ are relevant to models which attempt to establish a direct relation
between the critical temperature $T_{\rm c}$ and the strength of the $(x^2\!-\!y^2)\!-\!z^2$
coupling.
For La$_2$CuO$_4$, in particular, the  density-functional estimate used as input parameter
in such models \cite{CuO2_z2Tc_sakaki10} is about 0.5 eV smaller than our result.
Consequently, the difference we find between the $d_{x^2-y^2}$--$d_{z^2}$ splittings 
in La$_2$CuO$_4$ and HgBa$_2$CuO$_4$ is less spectacular as compared to the value 
reported by Sakakibara {\it et al.} \cite{CuO2_z2Tc_sakaki10}, suggesting a reevaluation of the analysis in Ref.~\cite{CuO2_z2Tc_sakaki10}.

\section{Methods}

The first step in our study is a restricted Hartree-Fock (RHF) calculation for the
ground-state configuration of the periodic system.
The RHF calculations are performed with the {\sc crystal} package \cite{crystal}.
We employed experimental lattice parameters
\cite{CuO2_rixsdd_10,str_LCO_longo73,str_HBCO_wagner93,str_YBCO_katano87,str_CCOC_argyriou95,str_SCOC_miller90}
and Gaussian-type atomic basis sets, i.e., triple-zeta basis sets from the {\sc crystal} library
for Cu, O, and Cl plus basis sets of either double-zeta or triple-zeta quality for the
other species.
Post Hartree-Fock many-body calculations are subsequently carried out on finite clusters,
which are sufficient because of the local character of the correlation hole.
They consist of five CuO$_4$ plaquettes, i.e., a ``central'' CuO$_4$ unit plus the
four NN plaquettes.
When present, the apical ligands, oxygen or chlorine, are incorporated as well in the
finite cluster $\mathcal{C}$.
Additionally, the finite cluster $\mathcal{C}$ includes in each case the NN closed-shell
metal ions around the ``central'' Cu site.
In La$_2$CuO$_4$, for example, there are ten La$^{3+}$ NN's.
In YBa$_2$Cu$_3$O$_6$, there are one Cu$^{1+}$ $3d^{10}$, four Y$^{3+}$, and four Ba$^{2+}$ NN's.

The orbital basis entering the post Hartree-Fock correlation treatment is a set of
projected RHF Wannier functions:
localized Wannier orbitals (WO's) are first obtained with the Wannier-Boys localization
module \cite{zicovich_01} of the {\sc crystal} package and subsequently projected
onto the set of Gaussian basis functions associated with the atomic sites
of $\mathcal{C}$ \cite{LaCoO_hozoi_08}.
Moreover, the RHF data is used to generate an effective embedding potential for the
five-plaquette fragment $\mathcal{C}$.
This potential is obtained from the Fock operator in the RHF calculation
\cite{LaCoO_hozoi_08} and models the surroundings of the finite cluster, i.e., the
remaining of the crystalline lattice.

The central CuO$_4$ plaquette and the four NN Cu sites form the active region of the
cluster, which we denote as $\mathcal{C}_A$.
The other ions in $\mathcal{C}$, i.e., each ligand coordination cage around the
four Cu NN's and the NN closed-shell metal ions, form a buffer region $\mathcal{C}_B$
whose role is to ensure an accurate description of the tails of the WO's centered in
the active part $\mathcal{C}_A$ \cite{LaCoO_hozoi_08}.
For our choice of $\mathcal{C}_B$, the norms of the projected WO's centered within
the active region $\mathcal{C}_A$ are not lower than $99.5\%$ of the original
crystal WO's.
While the occupied WO's in the buffer zone are kept frozen, all valence orbitals
centered at O and Cu sites in $\mathcal{C}_A$ (and their tails in $\mathcal{C}_B$) are
further reoptimized in multiconfiguration CASSCF calculations. 
In the latter, the ground-state wavefunction and the lowest four crystal-field excited 
states at the central Cu site are computed simultaneously in a state-averaged multiroot 
calculation \cite{QC_book_00}.
The $d$-level splittings at the central Cu site are finally obtained at the CASSCF+SDCI
level of theory as the relative energies of the crystal-field excited states.
The virtual orbital space in the multireference SDCI calculation cannot be presently
restricted just to the $\mathcal{C}_A$ region.
It thus includes virtual orbitals in both $\mathcal{C}_A$ and $\mathcal{C}_B$, which
leads to very large SDCI expansions, $\sim\!10^{9}$ Slater determinants for a FM
configuration.
For this reason, we restrict the CASSCF+SDCI calculations to FM allignment of the Cu
$d$ spins. 

The effective embedding potential is added to the one-electron Hamiltonian with the
help of the {\sc crystal-molpro} interface program \cite{crystal_molpro_int}.
Although the WO's at the atomic sites of $\mathcal{C}$ are derived for each of the
compounds discussed here by periodic RHF calculations for the Cu $3d^9$ electron
configuration, the embedding potentials are obtained by replacing the Cu$^{2+}$ $3d^9$
ions by closed-shell Zn$^{2+}$ $3d^{10}$ species.
This is a  good approximation for the farther $3d$-metal sites, as the comparison between
our results and RIXS data shows.
An extension of our embedding scheme toward the construction of open-shell embeddings
is planned for the near future.

%% \newpage

\section{Acknowledgements}

We thank V. Bisogni, L. Braicovich, G. Ghiringhelli, K. Wohlfeld, and V. Yushankhai
for fruitful discussions.
L.~H. acknowledges financial suport from the German Research Foundation (Deutsche Forschungsgemeinschaft,
DFG).

\section{Author Contributions}

L.H., P.F., and J.v.d.B. wrote the main manuscript text, L.H. and L.S. prepared figure 1.
All authors reviewed the manuscript.
%% All authors contributed to the main manuscript text. 

\section{Additional Information}

Competing financial interests: Authors have no Competing Financial Interests. 
%% The authors declare no competing financial interests.

\newpage

\section{Figure Legends}

{\bf Figure 1.\,}
Relative energy of the Cu $z^2$ level as function of the distance $h$ between
the Cu and apical ligands in different cuprates.
There are two apical O sites in La$_2$CuO$_4$ and HgBa$_2$CuO$_4$, two apical Cl
sites in Ca$_2$CuO$_2$Cl$_2$ and Sr$_2$CuO$_2$Cl$_2$, and one apical O ion in
YBa$_2$Cu$_3$O$_6$.
In CaCuO$_2$ the apical ligands are absent.

\newpage

%% TABLE 1
\begin{table}[!t]
\caption{
CASSCF+SDCI versus RIXS results for the Cu $d$-level splittings in La$_2$CuO$_4$,
Sr$_2$CuO$_2$Cl$_2$, and CaCuO$_2$ (eV).
The ground-state Cu $t_{2g}^6d_{z^2}^2d_{x^2-y^2}^{1}$ configuration is taken as reference.
A $2J$ term was here substracted from each of the RIXS values reported in
Ref.~\cite{CuO2_rixsdd_10}, see text.
}
\begin{ruledtabular}
\begin{tabular}{llll}
Hole orbital &La$_2$CuO$_4$   &Sr$_2$CuO$_2$Cl$_2$  &CaCuO$_2$  \\
             &SDCI/RIXS       &SDCI/RIXS            &SDCI/RIXS  \\
\colrule
\\
$x^2-y^2$    &$0$             &$0$                  &$0$            \\
$z^2$        &$1.37$/$1.44$   &$1.75$/$1.71$        &$2.38$/$2.39$ \\
$xy$         &$1.43$/$1.54$   &$1.16$/$1.24$        &$1.36$/$1.38$ \\
$xz,yz$      &$1.78$/$1.86$   &$1.69$/$1.58$        &$2.02$/$1.69$ \\
\end{tabular}
\end{ruledtabular}
\end{table}

\newpage

%% TABLE 2
\begin{table}[!t]
\caption{Cu $d$-level energy splittings for La$_2$CuO$_4$, HgBa$_2$CuO$_4$,
and YBa$_2$Cu$_3$O$_6$ (eV).
CASSCF+SDCI calculations for 5-plaquette FM clusters.
}
\begin{ruledtabular}
\begin{tabular}{llll}
Hole orbital &La$_2$CuO$_4$ &HgBa$_2$CuO$_4$ &YBa$_2$Cu$_3$O$_6$ \\
\colrule
\\
$x^2-y^2$    &$0$           &$0$             &$0$                \\
$z^2$        &$1.37$        &$2.09$          &$1.47$             \\
$xy$         &$1.43$        &$1.32$          &$1.22$             \\
$xz,yz$      &$1.78$        &$1.89$          &$1.57$             \\
\end{tabular}
\end{ruledtabular}
\end{table}

 \

 \

 \

 \

 \

 \

 \

 \

 \

 \

 \

 \

 \

 \

 \

 \

 \

 \

 \

 \

\newpage

%% FIGURE 1
\begin{figure}[!t]
\includegraphics*[width=0.65\columnwidth]{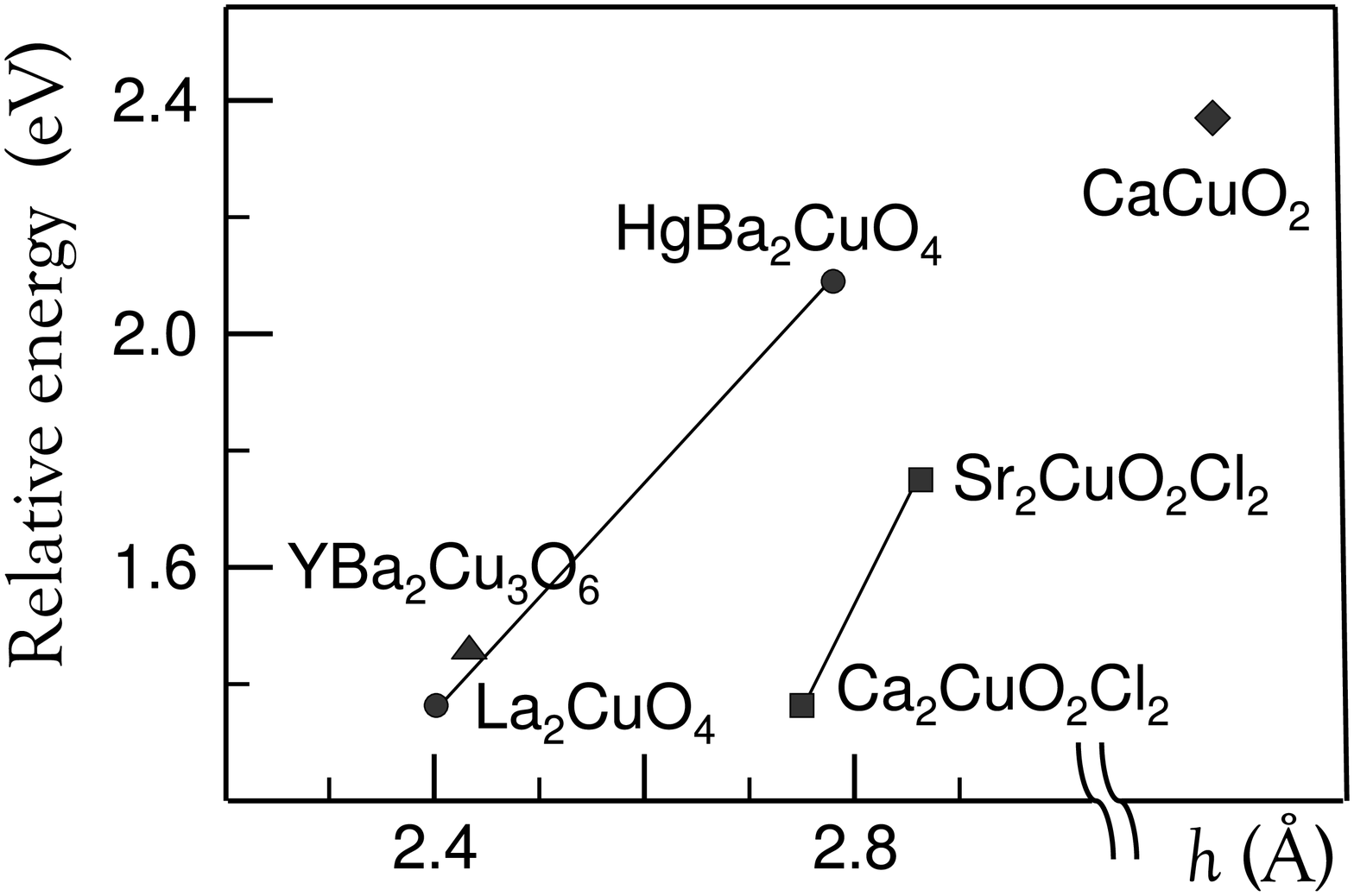}
\caption{
%% Relative energy of the Cu $z^2$ level as function of the distance $h$ between
%% the Cu and apical ligands in different cuprates.
%% There are two apical O sites in La$_2$CuO$_4$ and HgBa$_2$CuO$_4$, two apical Cl
%% sites in Ca$_2$CuO$_2$Cl$_2$ and Sr$_2$CuO$_2$Cl$_2$, one apical O ion in
%% YBa$_2$Cu$_3$O$_6$, and no apical ligands in CaCuO$_2$.
}
\end{figure}

%%%%%%%%%%%%%%
\end{document}